# Development of a non-wearable support robot capable of reproducing natural standing-up movements

Atsuya Kusui[1], Susumu Hirai[1,2], Asuka Takai[1]
[1] Osaka Municipal University, Graduate School of Engineering, Department of Mechanical Engineering
[2] ZEATEC co.,ltd.

Abstract   In order to reproduce natural standing-up motion, attention has been focused on the coordination of the movement of the assisting robot and the human. However, a number of previous non-wearable standing assistance devices have not been able to reproduce the natural trajectory. On the other hand, wearable devices are easier to coordinate with the body, but it is difficult to completely isolate mechanical and electrical hazards. Therefore, in this study, we aimed to develop a robot that can implement the natural standing-up motion of each user while achieving high coordination by integrating the features of the non-wearable type and the wearable type. This device adopts a four-link mechanism that is parallel to the segments and joint structure of the human body, and is designed to reproduce the S-shaped trajectory of the hip joint and the arc trajectory of the knee joint. The natural standing-up motion data of the subject was measured using a gyro sensor, and the link length was determined to drive the seat along the obtained trajectory. In addition, feedforward speed control by a stepping motor was introduced, and the trajectory reproducibility by the geometric constraint of the mechanism was evaluated. We designed and manufactured a standing motion assistance robot, and performed a load-bearing experiment with a weight fixed to the seat to evaluate the trajectory. As a result, the error in reproducing the trajectories of the hip and knee joints was within about 4% of the total movement of the seat, and a good agreement with the target trajectory was obtained. Furthermore, durability tests, thermal safety evaluations, and risk assessments were conducted to confirm the feasibility and safety of the device in indoor environments. The results of this research provide new guidelines for robot design that enable natural movement assistance according to the physical characteristics of each individual, and are expected to become a promising fundamental technology for future elderly support and rehabilitation applications.

## 1. Introduction

It has been reported that healthy adults stand up about 60 times per day on average(Dall and Kerr, 2010), and standing up from a chair is an essential basic movement for independent daily living activities(van Lummel et al., 2016, Duarte Wisnesky et al., 2020, Slaughter et al., 2015). When standing up, the lower limbs are subjected to a load that exceeds body weight regardless of the seat height (Yoshioka et al., 2014), which can make it difficult to stand up due to reduced motor function.

Against this background, many standing-up motion support devices have been developed to assist with care and physical functions (Nakamura and Saga, 2021, Chugo and Takase, 2009, Ruszala and Musa, 2005, Burnfield et al., 2012). For example, Chugo et al. have shown that a standing-up support device equipped with a mechanism to assist in leaning the upper body forward and extending the legs can reduce the load on the lower legs (Chugo and Takase, 2009).

However, in recent years, it has been pointed out that a design that drives only part of the body during standing can hinder the natural movement trajectory of other joints and impair the body's kinetic chain (Zhou et al., 2021). The kinetic chain here refers to the phenomenon in which changes in the movement or alignment of one part (segment) in a structure, such as a leg in which multiple joints are arranged in series spread through adjacent segments and affect the movement of the entire body. In fact, Ruszala et al. have reported that while existing standing motion support devices show a certain degree of effectiveness in rehabilitation, they may induce unnatural postures in users (Ruszala and Musa, 2005). Burnfield et al. also showed that even in stand-up assistance devices that aim to mimic the natural movement trajectory of healthy people, the reproducibility of the trajectory was insufficient (Burnfield et al., 2012).

As described above, with the progress of device development, standing assistance devices are now available in a variety of forms and with various assistance methods. In particular, the distinction between wearable and non-wearable types and classification according to user characteristics are useful for understanding the design concepts and expected usage scenarios of each device.

## 2. Analysis of the current paradigm of standing assistance devices

Standing assistance devices can be broadly classified into two types: "non-wearable" and "wearable" based on their structural form and the method of interaction with the user (Nakamura and Saga, 2021).

## 2.1 Characteristics of Wearable Systems

Wearable standing assistance devices are designed to be worn directly on the user's body, and assist the user's movements by directly applying force or torque generated by actuators (motors, pneumatic artificial muscles, etc.) to specific body parts or joints through structures arranged along the user's hip joints, knee joints, etc. (Shepherd and Rouse, 2017, Scaletta et al., 2016, Huo et al., 2022). Wearable devices include highly rigid exoskeletal types and flexible soft robot types, and a modular design has also been proposed that allows a flexible selection of body parts or joints to be supported depending on the user's physical condition and support needs (Noda et al., 2018)。

One of the main advantages of wearable systems is their ability to follow natural movements. Since it can be designed to fit the structure of the human body, it is expected to provide highly accurate assistance that matches natural movement trajectories (Sharma et al., 2024, Chen et al., 2017). In addition, since torque can be applied directly to joints, precise assist control can be realized (Chiaradia et al., 2024, Liao et al., 2022). Many wearable devices use motors as the main power source (Nakamura and Saga, 2021)。

On the other hand, wearable systems also have some limitations. First, there are issues with comfort and usability. Specifically, the harness and the fastenings to the body can cause a sense of pressure, blockage, and discomfort when using the system (Chae et al., 2021, Bae et al., 2022). In addition, the weight of the device itself may impose additional strain on the wearer (Zhou, 2024, Gumasing et al., 2023). Second, the effort of donning and doffing can be an issue (Rodríguez-Fernández et al., 2021, Reicherzer et al., 2024). Especially for elderly users and those with limited physical function, putting on and taking off the device can take time or require assistance. Third, there are alignment issues. It is difficult to achieve perfect alignment between the joint axes of a robot and the joint axes of the human body, and any misalignment can lead to unnatural forces, which can lead to unintended movements and discomfort (Näf et al., 2019, Stienen et al., 2009, Zanotto et al., 2015). Furthermore, ensuring safety is also an important issue. Since the structure involves close contact between the human body and the robot, it is difficult to completely isolate mechanical and electrical hazard sources, and essential safety must be ensured from the design stage (Fosch-Villaronga et al., 2023, Nasr et al., 2024, Bessler et al., 2021).

## 2.2 Characteristics of non-wearable systems

On the other hand, non-wearable standing assistance devices have a structure that supports and assists the body from the outside without being worn by the user. Non-wearable systems come in various forms, such as forward-facing types that support the trunk or receive support by holding a device installed in front of the user with both arms (Burnfield et al., 2012, Chugo and Takase, 2009, Xu et al., 2024), and rear-facing types that have knee support and seat elevation functions (Jeong et al., 2019, Lou et al., 2021a).

Representative devices include the lift-up chair, which assists with standing by tilting or raising the seat, and the standing frame, which provides external support for the trunk and lower limbs (Ruszala and Musa, 2005) These devices often employ four-bar linkages (Jeong et al., 2019) or simple pivot or telescopic mechanisms (Lou et al., 2021a, Xu et al., 2024), and actuators include linear actuators (Jeong et al., 2019, Xu et al., 2024, Takeda et al., 2018, Saint-Bauzel et al., 2009, Kuroda et al., 2022), motors (Rea et al., 2024, Cao et al., 2012, Matjačić et al., 2016) spring mechanisms (Nango et al., 2010), or hydraulic or pneumatic cylinders (Bulea and Triolo, 2012, Sogo et al., 2018, Bae and Moon, 2011, Suzuki et al., 2017, Kamnik and and Bajd, 2004).

The main advantage of non-wearable systems is their ease of use. They do not require a complicated wearing process and can be used immediately after the user approaches (Xu et al., 2024, Ruszala and Musa, 2005, Koumpouros et al., 2017). In addition, since there is no need to attach to the body directly, it is thought that the sense of pressure and discomfort associated with harnesses and other devices is less than that of wearable systems.

On the other hand, non-wearable devices also have some limitations. First, because of their indirect support structure, it may not be easy to provide precise support or a natural sensation compared to wearable devices (Lee et al., 2021). In addition, restrictions on the movement trajectory may be problematic, and the generated trajectory may be fixed or have a limited range of adjustment due to the device structure, so it may not fully match the natural movement of individual users (Purwar et al., 2023, Lou et al., 2021b). In particular, it has been pointed out that it is difficult to reproduce natural knee movement when using fixed knee pads for standing-up support (Fray et al., 2019). In addition, some devices using motors or linear actuators do not fully take into account the natural standing trajectory, and some reports only claim that they are close to natural (Matjačić et al., 2016). Stability can also be an issue; as the user is not physically restrained from the device, users may have difficulty maintaining stability during movement without the use of a harness, especially those with weak trunks or lower limbs (Fray et al., 2019, Jeyasurya et al., 2013).

## 2.3 Trade-off between wearable and non-wearable systems

As shown above, wearable and non-wearable standing assistance devices each have their own advantages and limitations, and there is a fundamental trade-off between the two. While wearable devices enable highly accurate assist control and personalized assistance, they also have issues such as comfort and the complexity of wearing them. Such discomfort and operational constraints have created a demand for non-wearable solutions that aim for simpler and more comfortable use. On the other hand, non-wearable devices have the limitation that, due to their structure, they tend only to provide indirect support and tend to have uniform movement trajectories. This leaves challenges in achieving natural and highly personalized exercise support.

From this perspective, the choice of which system paradigm to adopt suggests a need to balance the possibility of highly individualized assistance (wearable) with the reduction of the burden on users and ease of integration into daily life (non-wearable). Additionally, the safety design considerations are significantly different between the two. In particular, wearable types, which have a high degree of contact with the human body, require stricter measures to deal with physical and electrical hazards, while the main issues for non-wearable types are stable support and reducing the risk of falling.

Given this background, in recent years, there has been growing interest in hybrid approaches that are based on non-wearable technologies while partially incorporating the functional features of wearable technologies. In other words, the development of a feature-integrated system that combines the advantages of both is gaining attention as a new design guideline.

## 2.4 Previous research and issues regarding the hybrid approach

In recent years, there has been an increase in research into non-wearable standing support devices using advanced link mechanisms with the aim of reproducing natural human body trajectories. These are positioned as hybrid approaches that aim to achieve the natural movement tracking that wearable devices excel at while being non-wearable. Each study is developed based on different concerns and design concepts.

For example, Das et al. proposed a low-cost, compact device that can reproduce a natural axillary trajectory, focusing on the fact that many existing devices force users to follow unnatural movement trajectories (Das et al., 2022). Zhou et al. raised the issue that designs that drive only part of the body hinder the natural movement of other joints and reduce human-machine coordination, and attempted to solve this problem through a design based on the motion data of many people (Zhou et al., 2021) Purwar et al. (Purwar et al., 2023) proposed a device that emphasizes the reproduction of a natural hip joint trajectory from a rehabilitation perspective, preventing the establishment of incorrect compensatory movements and encouraging the relearning of correct movements (Purwar et al., 2023).

These studies vary greatly in their approaches depending on the design elements they emphasize. Das et al. achieved a four-position passage of the axilla using a four-link design based on Burmester theory, achieving high trajectory reproduction accuracy. However, the operation time was long at 15 seconds, and the configuration prioritized safety and accuracy (Das et al., 2022). Zhou et al. generated natural whole-body trajectories from the average trajectories of 26 subjects using spline interpolation, and confirmed the validity of the method through experiments with three subjects. In terms of position reproduction, accuracy varied depending on the height of the subject. The standing-up operation time was assumed to be 1.6 seconds, and emphasis was placed on speed by driving at a speed of 25 mm/s or less (Zhou et al., 2021). Purwar et al. used natural motion extraction using Kinect and smooth S-shaped trajectory generation using $C^2$ interpolation to design a six-link mechanism to reproduce the natural movement of the hip joint while moving 19 inches vertically within 30 seconds (Purwar et al., 2023).

However, all of these studies have design assumptions and limitations. The link length of the device by Das et al. is fixed (Das et al., 2022), and it cannot accommodate differences in body size. The device by Zhou et al. also showed variation in trajectory accuracy depending on the subject's height. However, the variability of the link length was not described, and the versatility of the structure remains an issue (Zhou et al., 2021). The device by Purwar et al. is said to be compatible with a range of body sizes from 5 to 6 feet and up to 158 kg(Purwar et al., 2023), but no quantitative evaluation of the reproducibility of the trajectory has been performed.

If the structure remains incapable of accommodating the different heights and movement characteristics of each user, it could hinder the natural standing movement and lead to the development of inappropriate compensatory movements. Therefore, it is necessary to introduce design methods that allow dynamic adjustment of the mechanism dimensions and trajectory according to individual differences, and to verify the accuracy of movement and mechanical characteristics quantitatively.

## 2.5 Standing Assist Device's System and Issues

Standing assistance devices available in Japan have characteristics that span multiple categories in the international standard ISO

9999, "Assistive devices for persons with disabilities -- Classification and terminology," due to their design concept and functional characteristics. This includes medical devices, mobility aids, furniture, etc. ISO 9999 was first established in 1992 and has been revised several times up to the fourth edition in 2007, providing a framework for systematically classifying available devices.

Based on this standard, standing assistance devices are broadly categorized into the following four types: 1. passive/mechanical assistive devices (e.g., quad canes, handrails, non-powered seating aids), 2. powered furniture types (e.g., powered lift chairs, etc.), 3. standing lifts (manual/powered, forward/backward, etc.), and 4. robotic technology devices (including wearable and non-wearable types). In particular, for standing lifts, differences in support methods from the front or rear and drive methods are taken into consideration depending on the subject's level of physical function.

These devices may also be classified based not only on their form but also on the degree of independence they provide to their users. Independence levels are roughly categorized into six stages: Stage 1: completely independent to requires supervision; Stages 2-3: supervision to partial assistance required; Stage 4: difficulty standing up on one's own and requires assistance such as a standing lift; Stage 5-6: full assistance required, use of a sling lift recommended.

However, in the current market, there is a tendency for devices such as standing lifts to be introduced even for users who are at a relatively high level of independence, such as at stage 2. Such excessive support not only has the potential to hinder the natural movement trajectory of other joints by being designed to drive only a part of the body but also has the problem of depriving the user of the ability to utilize their remaining functions (Fray et al., 2019).

### 2.6 Design concept and device overview of this research

This study aims to develop a novel sit-to-stand assistive robot designed to move in coordination with the user, based on a design philosophy that emphasizes the reproduction of natural motion tailored to individual differences. The objective is to enable users to engage their residual motor functions to the fullest extent by facilitating body movement patterns that closely resemble those in unassisted conditions. This design principle is consistent with the concept of active participation (Fray et al., 2019), which has been emphasized in the field of rehabilitation in recent years, and is intended to utilize and promote the user's physical functions, rather than simply acting as a substitute for the user's movements. In particular, this study focused on the importance of coordination between a person's natural standing-up movement and the movement of the assistive robot.

Recent studies have increasingly focused on the extent to which non-wearable sit-to-stand assistive devices can accurately reproduce natural movement patterns.(Das et al., 2022, Zhou et al., 2021, Purwar et al., 2023) . However, current non-wearable devices have difficulty fully reproducing a person's natural standing trajectory, and there are limitations in realizing precise coordinated motion. On the other hand, although wearable devices adhere closely to the body and are highly cooperative, there is a large trade-off between the sense of mechanical restraint and safety, and they are often not suitable for everyday use.

Therefore, in this study, we adopted a hybrid approach that integrates the features of wearable and non-wearable devices, and designed it to take advantage of the strengths of both while compensating for their weaknesses. The wearable elements incorporate the following: a link design based on the human body structure, an adjustment mechanism according to the length of the body segments, and precise position and speed control of each joint. On the other hand, non-wearable elements include support that does not require fastenings to the body, easy and quick start to use, and safe support through indirect torque application. By combining these, this study aims to realize natural support as a "robot that stands up together with the person."

This device is primarily targeted at users who have mild difficulty standing up (Stage 2 level of independence: supervision to partial assistance). It is designed for users who are partially able to stand up on their own but require assistance, such as older people and patients undergoing rehabilitation. Additionally, taking into consideration adaptability to individual differences, it is designed to accommodate physical conditions ranging from 140 to 190 cm in height and weight of approximately 80 kg (the design rationale will be explained in detail in the next chapter).

In this report, we used a uniquely developed standing assistance robot to verify the stability of movement and trajectory reproducibility through experiments in which the movement speed and weight conditions were changed.

### 3. Design of a standing-up support robot

Many existing nursing and welfare devices operate according to a preset target trajectory. However, when targeting users with various physical characteristics and reduced physical functions, it is considered desirable for the movement trajectory guided by the support to be adapted to the trajectory that minimizes the load on the body for each individual. Our previous research has shown that it is possible to derive the optimal standing-up movement trajectory that minimizes the physical load for each healthy adult (Takai et al., 2013). In this report, we aim to implement this optimal trajectory as the target movement of the robot and design a non-wearable support robot that assists the standing-up movement from a chair based on the following specifications.

### 3.1 Motion reproduction based on subject data and design principles of link structure

In this study, we designed a target trajectory using the standing-up movement of one subject (a man in his 20s) as an example. A gyro sensor was used for measurement, and the angle changes of the ankle joint, knee joint, and hip joint were obtained in a time series in the sagittal plane. During the measurements, we instructed the subjects to avoid hunching over and to keep their upper body in a straight line, and to keep the soles of their feet from coming off the floor, in order to ensure consistency with the analytical model of the human body. The subject set the foot position and standing speed according to the conditions that he felt were natural and easy to stand up. Informed consent was obtained from the subject before the experiment.

The subjects stood up from a chair with a seat height of 35.4 cm three times to ensure reproducibility. The obtained angular velocity data was then substituted into the equation of motion to calculate the physical load, and time-series data for the optimal ankle, knee, and hip joint angles suitable for the subject was derived based on a genetic algorithm, a type of evolutionary computation method (Figure 1A). The human body model was expressed as a two-dimensional link model of three rigid bodies in the sagittal plane (Takai et al., 2013) ). The ankle joints are fixed points, and the mass is assumed to be concentrated at the center of gravity of each rigid body. The length and weight of each body segment of the subject were measured, and the mass ratio of each segment and the position of each center of gravity were determined based on the handbook on human body size. For details, please refer to the literature (Takai et al., 2013). This analysis obtained a natural time-varying trajectory in which the hip joint center drew an S-shaped trajectory and the knee joint center drew an arc trajectory (Figure 1B). This S-shaped trajectory is also observed during natural standing movements (Purwar et al., 2023). In the next section, we describe a four-bar linkage mechanism designed to ensure that the centers of the hip and knee joints of a subject seated on the seat accurately follow these target trajectories in time.

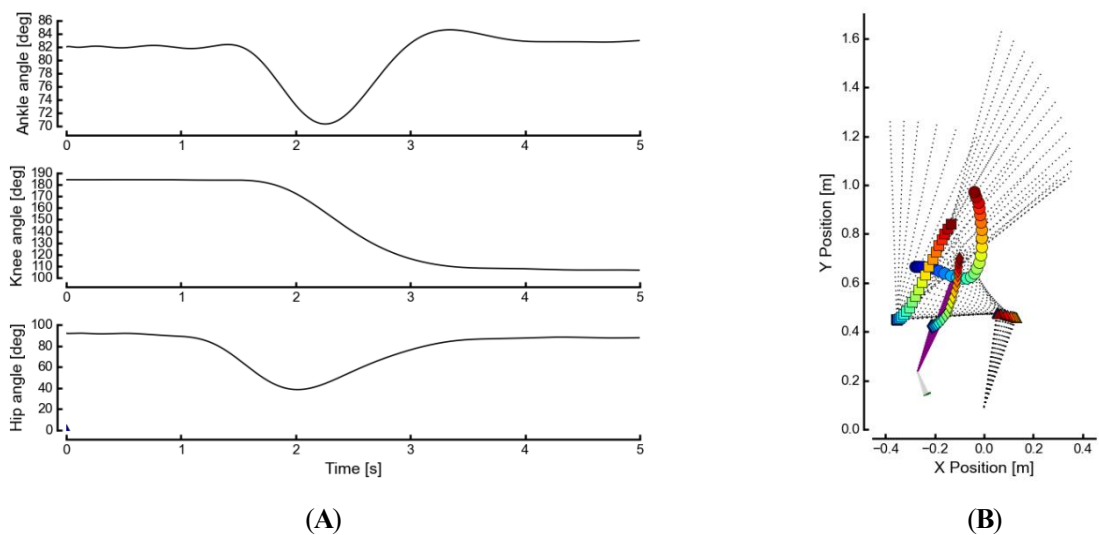

(A)                              (B)

**Fig. 1: Natural standing movement trajectory.**
(**A**) Time series data of joint angles optimized for an individual obtained by a genetic algorithm. (**B**) Movement trajectory in the sagittal plane was reconstructed based on the obtained angle data. Each part of the body is drawn with a black dotted line, and the knee joint, hip joint, and center of gravity of the body are indicated by △, □, and ○ markers, respectively. The change over time from sitting to standing is represented by a gradient from blue to red. The purple solid line represents the main linear actuator, with a diamond-shaped marker attached to the top end and connected to the seat (seat not shown). The green solid line shows the contraction trajectory of the sub-linear actuator.

### 3.2 Design principle of link structure

The four-bar link mechanism of this device shown in Figure 2 is composed of four links, links $A, B, C,$ and $D$. Links $A$ and $C$ are both basic structural parts of the device, and their lengths are constant and independent of the subject. On the other hand, link $D$ is a variable length link that is set according to the body dimensions of the subject, and link $B$ is a variable length link that is determined from moment to moment by motion calculations that will be described later.

The four-bar link mechanism used in this device is a planar link structure that is widely used in industrial robots and movable structures, and is composed of four rigid links and four rotational joints. This mechanism has the characteristic that one input link can drive other links, and has an excellent degree of freedom control and structural rigidity.

Furthermore, in this configuration, two triangular structures can be formed inside due to the geometric arrangement of the links. One is a triangle consisting of links $A$ and $B$ and the diagonal virtual link $L$, and the other is a triangle consisting of links $C$ and $D$ as well as $L$. In these triangles, if the length of each link is known, the corresponding angle is uniquely determined by the theorems of Euclidean geometry. By having such a triangular structure inside, the geometric constraints on the entire mechanism are strengthened, and resistance to structural deformation is increased. Therefore, the four-bar link mechanism is an effective design choice for standing assistance robots, which require high structural stability and precision in motion reproducibility.

The behavior of the four-bar link mechanism is shown in Fig. 3. After forming the subject's sitting posture from a chair-type configuration (i.e., the lower leg links are vertical and the seat is horizontal), the device operates along the optimal joint angle trajectory obtained in Section 3.1.

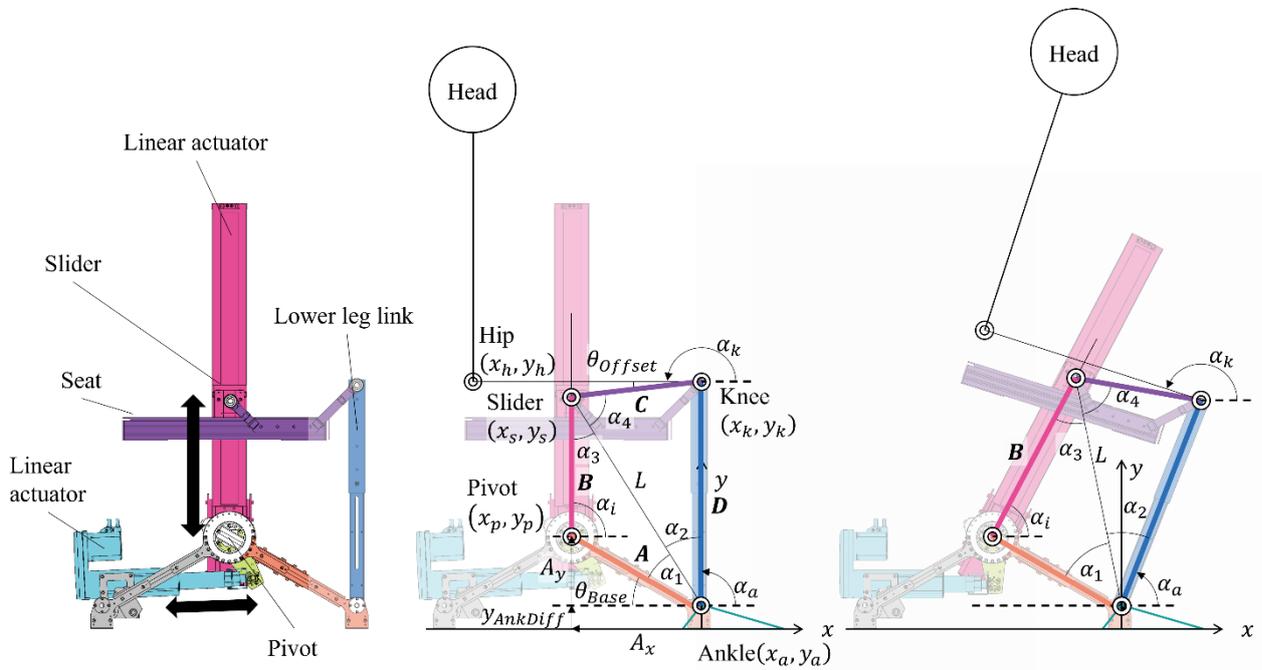

**Fig. 2: Configuration of the four-bar link mechanism used in this device.**
It is composed of four rigid links, links A, B, C, and D, and four rotary joints. Links A and C have fixed lengths, link D is set according to the subject's physique, and link B is a passive link that changes during movement.

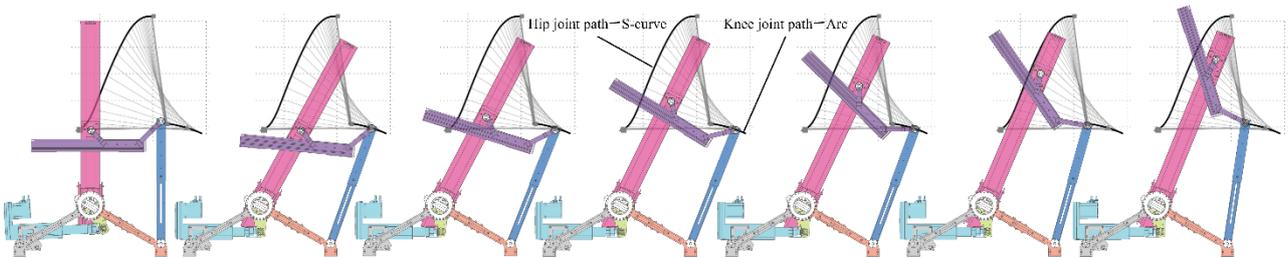

**Fig. 3: Behavior of the four-bar link mechanism.**
The subject's sitting posture is formed from the initial posture (left end) with the seat surface horizontal and the lower leg link vertical, and then each joint moves in conjunction with the other along the target trajectory obtained in Section 3.1.

### 3.2.1 Calculation of the length and angle of the main actuator

In this study, the length $L_{M,k}$ and rotation angle $\alpha_{i,k}$ of the main actuator at each time step were calculated discretely based on the joint angle data at each time step $k$ obtained from the subject and the geometric configuration of the link mechanism. All positional relationships are defined in a Cartesian coordinate system in the sagittal plane.

The length of the main actuator, $L_{M,k}$, is expressed as the Euclidean distance between the actuator fixed point $P = (x_p, y_p)$ and the linear actuator slider position point $S_k = (x_{s,k}, y_{s,k})$, as $L_{M,k} = \sqrt{X_k^2 + Y_k^2}$. Here, the coordinate differences $X_k$ and $Y_k$ are expressed as follows:

$$X_k = x_a + D \cos \alpha_{a,k} + C \cos(\alpha_{k,k} + \theta_{Offset}) - x_p$$
$$X_k = y_a + D \sin \alpha_{a,k} + C \sin(\alpha_{k,k} + \theta_{Offset}) - y_p$$

Here, $(x_a, y_a)$ are the coordinates of the ankle joint center, $D$ is the lower leg link length, $\alpha_{a,k}$ is the ankle joint angle, $\alpha_{k,k}$ is the knee joint angle, $C$ is the link length from the knee joint to the center of the slider, and $\theta_{Offset}$ is the offset angle between the link $C$ and the thigh axis direction.

The rotational angle $\alpha_{i,k}$ of the main actuator can be calculated using the following conditional arctangent function based on the coordinate difference between fixed point $P$ and slider position point $S_k$.

$$\alpha_{i,k} = \begin{cases} \pi + \tan^{-1}\left(\frac{Y_k}{X_k}\right), & (X_k < 0) \\ \tan^{-1}\left(\frac{Y_k}{X_k}\right), & (X_k > 0) \\ \frac{\pi}{2}, & (X_k = 0) \end{cases}$$

This uniquely determines not only the actuator rotation angle, but also the other link positions and angles.

Based on the joint angle data $\alpha_{a,k}$ and $\alpha_{k,k}$ obtained from the subject, the length of each link and the angle between each link that makes up the four-bar link mechanism were calculated sequentially at each time step $k$ to optimize the structure (Fig. 1(B)). In particular, the required vertical thrust is greatest just before the buttocks leave the seat. For this reason, in this design, the link arrangement and the actuator mounting angle were optimized so that the user's center of gravity is close to the thrust axis of the linear actuator at that moment (the area where the marker color changes from light blue to yellow-green in the figure). This is configured to maximize the lifting effect. Note that this design is based on a static mechanical model and does not take into account dynamic factors such as inertia and friction. This minimizes thrust loss and enables efficient lifting with little assist force.

The upper part of Fig. 5 shows the time change in the length of the main actuator at each time step.

### 3.2.2 Calculation of the length of the sub actuator

As shown in Figure 4, the length $L_{S,k}$ of the sub-actuator at each time step $k$ is calculated by geometrically defining the arm tip position point $T_k = (x_{T,k}, y_{T,k})$ with the rotation center point $P = (x_p, y_p)$ of the main actuator as the reference point. First, the coordinate $T_k$ of the arm tip is expressed as follows using the rotation angle $\alpha_{i,k}$ of the main actuator:

$$x_{T,k} = x_p + L_L \cos \alpha_{i,k}$$
$$y_{T,k} = y_p - L_L \sin \alpha_{i,k}$$

Here, $L_L = \sqrt{L_{Lx}^2 + L_{Ly}^2}$ is the link length from the rotation axis to the arm tip, and is defined as the following link vector components.

$$L_{Lx} = 0.050[\text{m}], \quad L_{Ly} = 0.0866[\text{m}]$$

Note that the link length is $L_L = 0.100$ [m].

Next, the distance from point $P$ to point $T_k$ is defined as the length of the sub-actuator $L_{S,k}$. The link attachment reference point at the position of point $P$ is geometrically expressed as $(x_p + L_{Lx}, y_p - L_{Ly})$, so $L_{S,k}$ can finally be calculated by the following formula:

$$L_{S,k} = \sqrt{\left(x_{T,k} - (x_p + L_{Lx})\right)^2 + \left(y_{T,k} - (y_p + L_{Ly})\right)^2}$$

In this way, the length of the sub-actuator is geometrically uniquely defined based on the relative positional relationship between the position $T_k$ of the arm tip and the rotation center point $P$.

The reproduction of the assistive motion in this device is achieved by a design method that combines motion data based on actual measurements and the geometric characteristics of the link structure. This ensures consistency between the structure and the motion, and is expected to function as a highly reliable assistive mechanism. The lower part of Figure 5 shows the time change in the length of the sub-actuator at each time step.

### 3.2.3 Actuator pulse frequency calculation method

In this device, the speed is calculated based on the change in length of the main and sub-actuators, and the pulse frequency required for control is calculated based on this. Here, the driving distance of each actuator is converted to micrometers, and a command signal to the stepping motor is generated based on the moving distance per pulse.

For example, first convert the length of the main actuator $L_{M,k}$ from meters to micrometers.

$$L_{M,k} = L_{M,k} 10^6 \ [\mu m]$$

Next, the discrete change in length, that is, the speed (difference approximation), is defined by the following equation:

$$\dot{L}_{M,k} = L_{M,k+1} - L_{M,k}$$

Here, when the moving distance per pulse is $\delta = 6 \ [\mu m/pulse]$, the required pulse frequency $f_{M,k}$ is given by the following equation.

$$f_{M,k} = \frac{\dot{L}_{M,k}}{\delta} \times 100$$

Here, "×100" is the time scaling assuming that one step is 0.01 seconds (i.e., a control period of 100 Hz).

Similarly, the pulse frequency for the length of the sub-actuator is calculated using the same method as above.

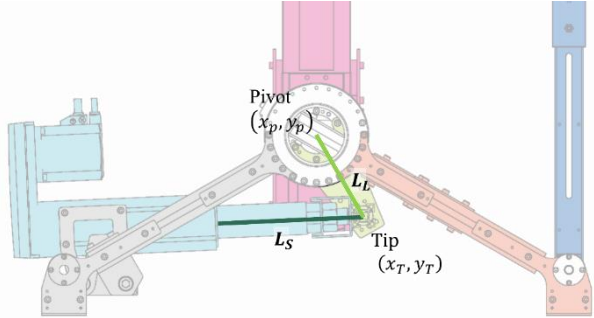

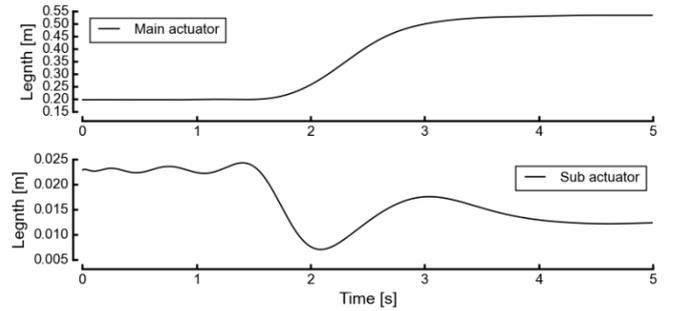

**Fig. 4: Geometric configuration used to calculate the sub-actuator length.** The arm tip point $T_k$ is defined from the rotation center point $P$ of the main actuator, and the length of the sub-actuator is the distance from that point to the reference attachment point. The link length $L_L$ is derived from the vector components.

**Fig. 5: Change in actuator length over time.** The changes in length of the main actuator (top) and sub-actuator (bottom) are calculated in response to the user's standing-up motion.

### 3.3 Design basis and target users based on subject data.

The design of this device was based on the human body size database of approximately 500 Japanese adult men and women surveyed by the National Institute of Advanced Industrial Science and Technology (AIST) in 1991–1992. The target data included young people (18–29 years old: 217 men, 204 women) and older people (60 years old or older: 50 men, 50 women) and included a wide range of dimensional information such as lower limb length, buttock width, and ankle height.

The $x$-distance from the ankle joint to the actuator's rotation center point $P$ was set to 270 mm based on the AIST seated buttock-lateral epicondyle distance (J7) and seated buttock-trochanter distance (J9). The $y$-distance was set to 145 mm based on the AIST seated popliteal height (I15), assuming a seat thickness of 40 mm. The above dimensions were fine-tuned and optimized so that the user's center of gravity would be close to the thrust axis of the linear actuator at the moment when the vertical thrust was maximum (from light blue to yellow-green in Figure 1B) just before buttock release (when the buttocks leave the seat) (see Supplementary Figure 1).

The ankle joint's $x$-direction position was set to 0 mm, and the $y$-direction (height) was designed to be 50 mm or more, based on the minimum value of the AIST lateral malleolus (M4), 52 mm, taking into account the wearing of shoes and sensors. This height can be changed to suit the user using the adjustment block depending on the shoe and height (see Section 3.4).

The lower leg link length $D$ is based on the value obtained by subtracting the AIST lateral malleolus (M4) from the AIST tibial superior border height (B25), and can be adjusted in stages in the range of 340 mm to 440 mm using pins in 50 mm increments. It can also be adjusted steplessly to suit the user's lower leg length.

The front-to-back seat width was also set by subtracting the average of the seated buttock-trochanter distance (J9) from the AIST seated buttock-popliteal distance (J6). Based on this distribution, it was designed to accommodate a wide range of body types.

The seat width was determined by taking into consideration the measurement data of seated buttock width: leg drop (J1) and seated buttock width: plantar support (J2) related to buttock width (seat width). In order to prevent the imbalance of the load due to lopsided sitting, a margin of 10 mm was provided on both sides of the seated buttocks, making the seat width 370 mm. For general chairs and wheelchairs, a margin of about +50 mm relative to the greater trochanter width is considered appropriate, and seat widths of 450 to 520 mm are often adopted.

As for shoulder width, the inside width of the main actuator is approximately 528 mm, so the target users are those whose shoulder width does not exceed this.

The link configuration above is based on the average lower leg length, and the design range of heights that can be accommodated is approximately from the 10th percentile of the elderly population (approximately 140 cm) to approximately 190 cm. On the other hand, short people below the 5th percentile of the elderly population and tall people over 190 cm may not be eligible. Based on static dynamics analysis, in these physique conditions, the linear actuator may not be long enough or may not have enough thrust. As a future expansion plan, it is possible to consider introducing an auxiliary mechanism such as a gas spring.

## 4. Configuration of the standing assistance device

The appearance of the standing assistance robot developed in this study is shown in Figure 6.

### 4.1 External dimensions and power supply specifications

The external dimensions of this robot are 676 mm wide, 893 mm deep, and 1025 mm high. The standing assistance unit weighs approximately 40 kg, and the control unit (hereafter referred to as the control tower) installed on the back weighs approximately 10 kg. The power supply is compatible with household AC 100 V, and the rated power consumption is 500 W or less.

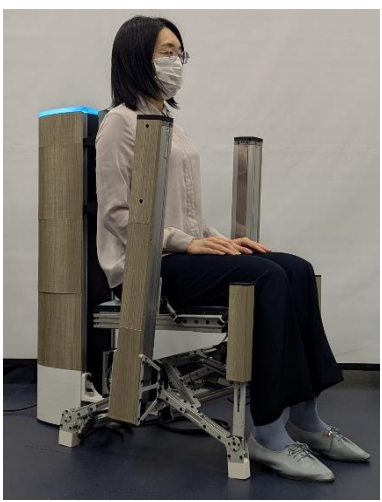

**Fig. 6: Appearance of the support robot.**

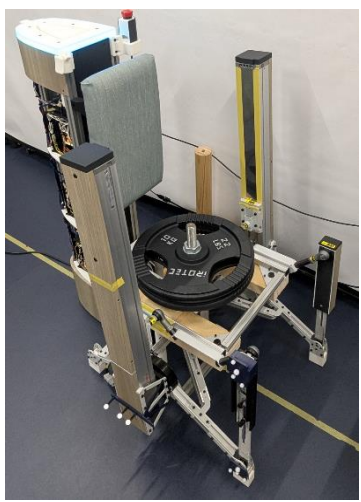

**Fig. 7: Appearance of the load fixing method using a wooden seat.**

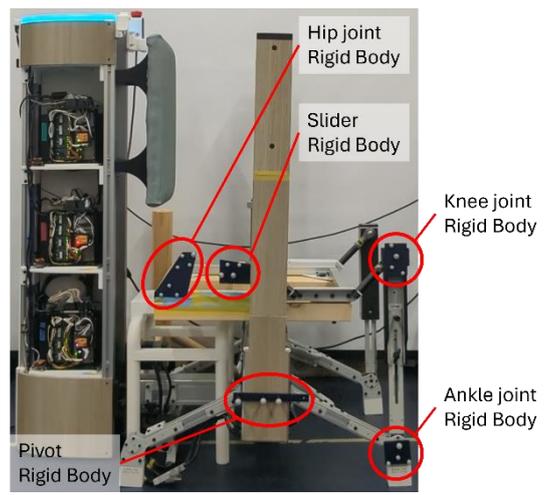

**Fig. 8: Marker placement for each joint.**

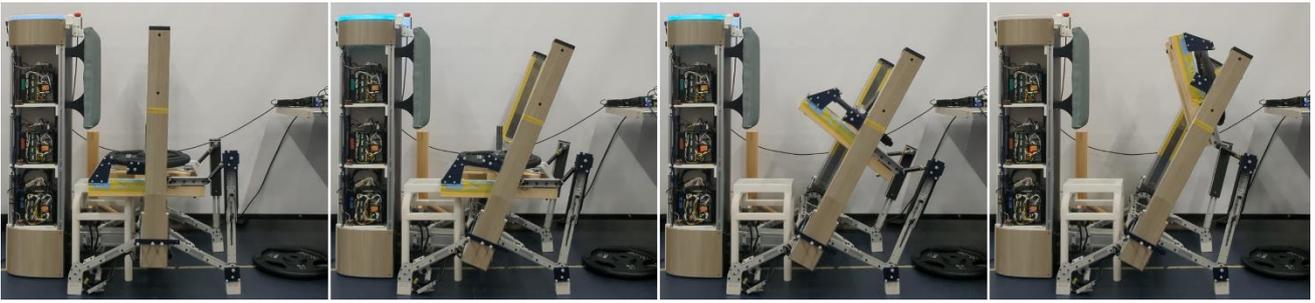

**Fig. 9: Evaluation experiment under a fixed load condition using a wooden seat.**
A 10 kg load is fixed onto the standing assistance device, and the trajectory is measured using a motion capture system.

### 4.2 Mechanism and Actuator

This robot is composed of a pair of main linear actuators, one on each side, which controls the seat in the up-down direction, and a sub-linear actuator, which controls the seat in the forward-backward direction. Both use motors (αStep AZ series) manufactured by Oriental Motor Co., Ltd., and are driven by a stepping motor and ball screw mechanism that converts rotational motion into linear motion. All three are equipped with AZM66AC motors, and the linear movement distance (lead length) per rotation is 6 mm.

The specifications of the main actuator are as follows: The dimensions of the slider body are a width of 74 mm and a height of 66.5 mm. The stroke is 500 mm. The minimum movement is 0.01 mm. The maximum payload is 60 kg (horizontal direction), the maximum thrust is 400 N, and the maximum pressing force is 500 N (at a speed of 25 mm/s or less). The holding force is also 400 N, the maximum speed is 400 mm/s, and the mass of the actuator is approximately 6.1 kg. At speeds below 200 mm/s, a maximum load of 30 kg can be transported, and the thrust decreases linearly when the speed reaches 200 to 400 mm/s.

The specifications of the sub-actuator are as follows. The dimensions of the slider body are a width of 60 mm and a height of 60 mm. The stroke is 100 mm, and the minimum movement is 0.01 mm. The maximum payload is 60 kg, the maximum thrust is 360 N, the pushing force is 500 N, and the holding force is also 360 N. The maximum speed is 300 mm/s, and the actuator mass is 3.0 kg. The thrust characteristic is maintained at 360 N from 0 to approximately 280 mm/s, and then drops sharply to around 300 mm/s.

Based on these specifications and the output limit of the main actuator, the maximum weight of the intended user of this device is estimated to be approximately 80 kg. The thrust specifications of the sub-actuator were determined by the following simple experiment, in addition to the static analysis results shown in the Supplementary Figure. Specifically, the design parameters were selected based on the minimum operating force required when a healthy person is seated on the wooden frame and the experimenter manually pulls it forward, so that this value would be approximately 70% of the rated output.

### 4.3 Mechanical structure and link mechanism

The movement of the seat in this device is controlled via a rotating arm and link mechanism (see Figure 4). The link driven by the sub-actuator rotates the rotating arm connected to the left and right main actuators on an orbit with a radius of 0.1 m, moving the seat along a specified orbit. A single-row deep groove ball bearing is placed at the center of rotation of this rotating arm to ensure high structural rigidity and smooth rotation characteristics.

The seat and main actuator are connected by a slider and an L-shaped rod end bearing (ball type), and the structure is designed to minimize the stress reaching the frame even if there is a slight misalignment in the operation synchronization between the left and right actuators. The range of motion is approximately ±20 degrees.

The lower leg link is connected to the seat via a rod end bearing at the connection point (knee joint position), and a small deep groove ball bearing is placed at the ankle joint to achieve smooth foot rotation. A spacer-type adjustment block made by a 3D printer is used to adjust the height of the ankle. The adjustment is designed based on the height from the lateral malleolus (ankle) to the ground, and in the initial state (spacer thickness ±0 mm), the ankle height is set to 50 mm from the ground. This design standard is based on statistical data on lateral malleolus height (M4) from the National Institute of Advanced Industrial Science and Technology (AIST), and since the minimum value when barefoot is 52 mm, it is determined that there will be no problems in actual use, even when taking into account the wearing of shoes and the placement of sensors on the soles of the feet. Since the maximum lateral malleolus height in the AIST data was 88 mm, four types of spacers (0 mm, +13 mm, +27 mm, +41 mm) in 13 mm increments were prepared to cover the range of approximately 50–90 mm for this device. This allows flexible adjustment according to the user's foot dimensions and ensures higher accuracy in matching the reproduced trajectory and the positional relationship of the body.

### 4.4 Covers

The surface of the seat is covered with waterproof chair fabric, and cushioning and load cells are embedded inside to enable sensing of the user's weight. The main actuator is covered with a 3D-printed cover, which is then covered with wood-grain wallpaper to enhance the chair's harmony in the indoor environment.

### 4.5 Operation method

The user sits on the seat in the center of the robot, aligns the position of the ankles and knees with the center of the robot's joints, and then presses the operation switch connected to the controller to start the standing-up motion. After standing up, the robot returns to a seated position by following the reverse operation sequence.

### 4.6 Safety features

This device is equipped with the following three types of safety mechanisms. 1. Detection of the difference in the amount of movement between the left and right actuators: A synchronization error is detected by software processing, and a control signal automatically stops the device from the main microcomputer. 2. Electrical emergency stop function: A physical button is installed that allows the user or experimenter to stop the device immediately. 3. Mechanical stopper: A structural limit is set to prevent overrotation and abnormal posture in the link mechanism. In addition, since the linear actuator installed in this device uses a stepping motor without a brake mechanism, the seat is designed to descend slowly under its own weight when the power supply is cut off.

### 4.7 Mechanical constraints and structural risks

The link mechanism used in this device has a mechanical limit in that it has a singular posture. Specifically, in a singular posture where the links that make up the robot are aligned in a straight line, structural ambiguity occurs in which the bending and extension direction of the knee joint cannot be uniquely determined only by the length and rotation angle of the main actuator. To address this issue, this device is designed to set appropriate limits on the control range and operation sequence to prevent it from reaching a singular position.

In addition, since the seat is raised and lowered by the left and right main actuators, if the synchronization of the actuators is lost, there is a risk that the seat will tilt or distort, and even lead to deformation of the entire frame. The following two-fold countermeasures are taken against this structural risk: 1. Rod end bearings ensure structural flexibility (stress relief). 2. Software detects synchronization errors and automatically stops the system. This ensures safety from both the mechanical and control perspectives.

### 4.8 Control System Configuration

The control tower contains an electronic circuit unit with a five-layer configuration. Each layer is configured as follows. 1st layer: Power supply circuit (24 V power supply for motor control and 5 V power supply for microcontroller). 2nd layer: Motor driver (Oriental Motor AZD series) for the left main actuator and homemade relay board. 3rd layer: Motor driver unit for the right main actuator. 4th layer: Motor driver unit for the sub-actuator. 5th layer: Bridge circuit for load cell signal and microcontroller for indicator control.

The main control microcontroller is M5Stack Core S3, and a total of five sub-microcontrollers (M5Stamp S3) are connected to it via I²C communication. Each sub-microcontroller plays the following roles: Motor control (left and right main and sub-actuators), acquisition and processing of load cell signals, and control of status display indicators.

## 5. Experiment

In this study, we used the standing assistance robot we designed and manufactured to carry out a load-bearing experiment with a weight fixed to the seat, and evaluated the reproducibility of the trajectory. In addition, we also carried out an evaluation experiment of durability and thermal safety (Section 6.2) and a risk assessment (Section 6.3).

### 5.1 Experimental Method

In order to quantitatively evaluate the reproducibility of the standing trajectory, we used a wooden seat to which a weight can be fixed, rather than a cushioned seat (Fig. 7). The three-dimensional motion of each part of the robot was captured using an optical motion capture system (OptiTrack, NaturalPoint Inc.). Markers were attached to the following joints and registered as rigid bodies (Fig. 8): ankle, knee, and greater trochanter (all robot parts equivalent to the joints of the human lower limb), driving joint (corresponding to point $P$), and slider position of the main actuator (point $S_k$). The trajectory in the sagittal plane was extracted from the acquired three-dimensional data, and the trajectory and operation time of each joint were calculated. The sampling frequency of the motion capture was 120 Hz.

The robot was loaded under the following three conditions (0kg, 10kg, 20kg), and 15 motion trials were performed for each condition (Figure 9). The maximum load of 20 kg was set based on the results of a preliminary experiment in which the seat vertical (perpendicular to the seat) was measured with a load cell when a healthy adult stood up with assistance. In the preliminary experiment, 11 healthy adults weighing 65.3 ± 10.4 kg participated. Although there was some variation depending on the subject and the motion situation, the load applied to the seat just before standing was generally within the range of 10 to 20 kg.

Two motion speed conditions (5 and 10) were set. Here, speed 10 is based on the ideal standing speed based on the joint angle trajectory derived by the optimization algorithm. Speed 5 is 50% of that speed and was used to verify the effect of speed change on assistance performance and motion stability.

Furthermore, to quantitatively evaluate the load on each actuator, evaluation was performed using stepping motor control software MEXE02 Ver.4 (manufactured by Oriental Motor). The AZD-A driver and a PC with MEXE02 Ver.4 installed were connected via USB, and the load rate of each motor was obtained using the waveform monitor function. In the measurement, recording was started using the rising edge of the control signal "MOVE" to 1 (ON) as a trigger, and data was obtained for 10 seconds, which corresponds to the rising movement of the seat. The "MOVE" signal is an output signal that indicates the operating state of the motor, and in this setting, it is configured to be ON while the motor is driving. The data sampling period was 10 ms (100 Hz), and recording was limited to the ascending motion; the load during the descending motion was not included. The measurement targets were the sub-actuator, the left main actuator, and the right main actuator, and measurements were performed independently five times for each.

### 5.2 Experimental analysis method

In this experiment, the position data of the rigid marker obtained by motion capture was used to extract the movement start time. First, the arctangent function was applied to the coordinate series at each time to calculate the rotation angle in the sagittal plane. The obtained angle series was corrected to a relative angle based on the initial angle, and then the first-order time derivative was taken to derive the angular velocity series. The time difference was approximated using the average time interval in the series based on a sampling rate of 120 Hz. To determine the start and end of the movement, a reference threshold was defined from the angular velocity of the first 24 samples (about 0.2 seconds) that were considered to be in a stationary state. Specifically, the threshold was set to the average value of the angular velocity during the stationary period plus twice its standard deviation. Note that at the end of the movement, the angular velocity data contains noise due to the vibration of the device, so the threshold was set using data smoothed by a moving average (window size: 24 samples), and the movement end time was searched for from the smoothed data. Using these, the start of a movement was determined when the angular velocity exceeded this threshold, and the end of the movement was determined when it fell below this threshold. Movement start threshold: 0.051 ± 0.009, movement stop threshold: 0.003 ± 0.01.

To evaluate the reproducibility of the trajectory, we quantitatively calculated the geometric similarity between the subject's joint trajectory and the robot's joint trajectory captured by motion capture. For each point on both trajectories, the Euclidean distance to the nearest point on the other trajectory was calculated, and the median of the Euclidean distances was used as the error index. This is an overall evaluation of the distance between each point on both trajectories and the nearest point, and has the characteristic of being less susceptible to deviations near the endpoints and outliers. In particular, the trajectories studied in this study may not have the same starting and ending points, and the overall similarity may be underestimated by "distance evaluation based on the worst case," such as the Frechet distance and the Hausdorff distance. Therefore, we determined that the median of the nearest distance used in this study is appropriate for stable evaluation of the degree of agreement of the overall trajectory shape.

## 6.　Experimental Results
### 6.1 Loading Experiment Results

The results of comparing the movement speeds of 5 and 10 under the load conditions of 0 kg and 10 kg are shown below. The 20 kg condition is shown in the Supplement Materials, but it was excluded from the comparison in this paper because the motor load rate exceeded 100% after the end of the movement by the MOVE signal and an error occurred before the experimenter finished the operation.

Figure 10 shows the target standing assistance trajectory (S-shaped trajectory for the hip joint and arc trajectory for the knee joint) and the corresponding measured trajectory of the robot plotted in the sagittal plane. The measured trajectories generally reproduced the target trajectories well, and it was confirmed that the hip joint drew an S-shaped trajectory and the knee joint drew an arc trajectory. It was observed that the trajectory variance over the 15 trials was slightly larger at speed 10 compared to speed 5.

Figure 11 shows the results of the Root Mean Square Error (RMSE) for each trial for the average trajectory of 15 trials. The knee joint had an average error of about 6 mm, and the hip joint had an average error of about 12 mm. The error trends for each condition are visualized by color coding. In particular, the RMSE was the smallest for both the knee and hip joints under the condition of speed 10 and load 10 kg, while the RMSE tended to be the largest under the condition of speed 10 and load 0 kg. This is thought to be due to the fact that the seat is made of wood and is lightweight, which increases vibration during high-speed operation and increases the trajectory variation. On the other hand, it is presumed that adding load suppresses the vibration of the seat, increasing the stability of the reproduction.

Figure 12 shows the evaluation results using the median of the nearest-neighbor distance from the target trajectory as an index. For the knee joint, the median distance was almost zero, and it was confirmed that the knee joint drew a stable arc trajectory with the ankle joint, which is the center of rotation, as the base axis. This means that the behavior is consistent regardless of the load or speed conditions. On the other hand, a distance error of about 15 mm was observed for the hip joint, and the error tended to become larger regardless of the speed, especially under the condition of a load of 10 kg. An error of about 20 mm was confirmed at the maximum, and it can be seen from the trajectory in Figure 10 that the trajectory is slightly shifted downward due to the effect of the weight of the load. Note that the total movement amount is about 400 mm in both the X and Y directions, so the error of 15 mm corresponds to about 4% of the total, and is considered to be within the acceptable range for practical use.

The operating time was 8.329 ± 0.178 seconds at 0 kg and speed 5, 8.156 ± 0.247 seconds at 10 kg and speed 5, 4.448 ± 0.067 seconds at 0 kg and speed 10, and 4.357 ± 0.038 seconds at 0 kg and speed 10. The increase time from 0 kg to 10 kg was -2.081% at speed 5 and -2.061% at speed 10, showing a decreasing trend.

Figure 13 shows the evaluation results of the motor load rate of each actuator. Under all operating conditions, the load rate on the left and right main actuators was a maximum of approximately 80% on average, remaining within the range that the system can operate. For the sub-actuators, except for a sharp transient peak, the maximum load rate was approximately 40% at a load of 0 kg and reached approximately 60% at a load of 10 kg. At 20 kg as shown in the Supplement Materials, the maximum value reached approximately 80%, suggesting a tendency for the load factor of the sub-actuator to increase roughly linearly with respect to the load.

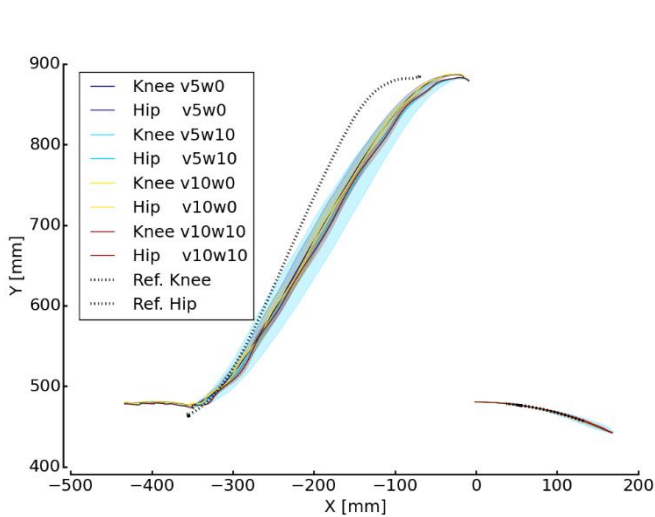 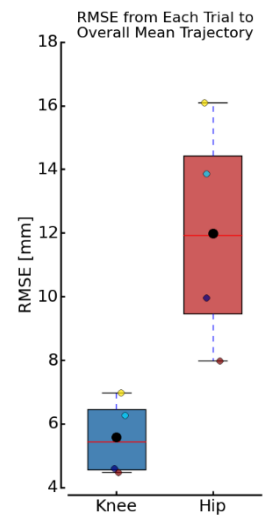 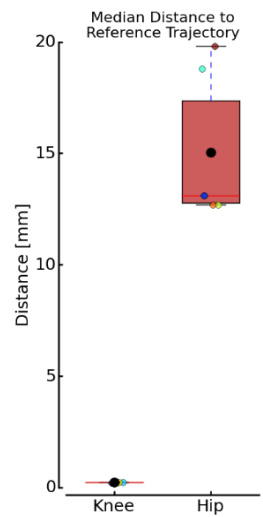

**Fig. 10: Comparison of hip and knee joint trajectories (in the sagittal plane).**
The measured trajectory of the robot under each condition is compared against the target trajectory (hip joint: S-shape, knee

**Fig. 11: Comparison of average trajectories.**
RMSE for each condition is shown to evaluate the

**Fig. 12: Evaluation of trajectory similarity based on nearest-neighbor distance.**

joint: circular arc).

condition dependency of the trajectory error. The median of the distance between each joint's points and the target trajectory was used as the evaluation index.

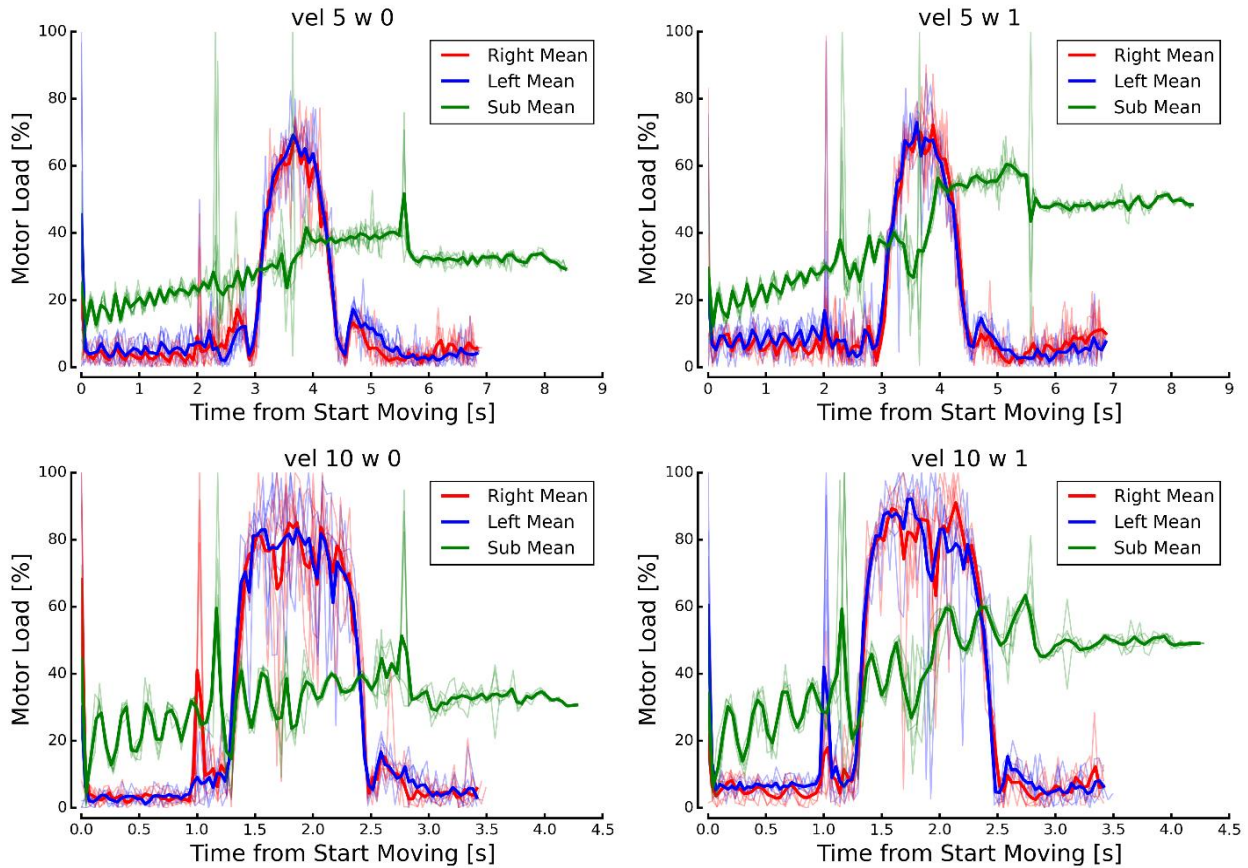

**Fig13: Time series changes in the motor load rate of each actuator.**
The motor load rates of the sub, left, and right actuators are shown.

### 6.2 Evaluation experiment of durability and thermal safety.

To evaluate the durability and thermal safety of the device, 150 consecutive lifting and lowering operations were performed with a load of 10 kg applied to the seat. The speed was set to 5, and the continuous operation was performed without any rest periods. The entire test took approximately 1 hour. These conditions were based on conditions assuming frequent use in indoor environments, and the expected locations of use include hospitals, welfare facilities, ordinary homes, public transportation (train seats), and seats in theaters and movie theaters.

The frequency of use in such environments was estimated to be approximately 30 to 200 times per day. For example, in a case where multiple users take turns using the device, this would be approximately 30 times per day (17 hours) at once every 30 minutes, or, considering the boarding and disembarking operations on public transportation, approximately 100 times per day at once every 10 minutes. The 150 consecutive operations performed this time are strong enough to cover these conditions, and are deemed to be appropriate verification conditions for confirming durability.

During the test, the temperature inside the control unit fluctuated between 28°C and 33°C, with no abnormal temperature rise observed. A thermally stable state was maintained, confirming the safety of the heat dissipation design.

Furthermore, in a separate preliminary experiential demonstration, approximately 30 subjects were repeatedly assisted in standing up, but there was no loosening or damage to the seat structure or fixings (especially the screws). These results demonstrate that the device has the mechanical reliability and thermal stability to withstand repeated use in a typical indoor environment.

### 6.3 Risk assessment.

The safety of this device was evaluated based on the risk assessment sheet recommended by the Project to Promote the Development and Introduction of Robotic Devices for Nursing Care. The evaluation method used was to calculate a risk estimate for each risk factor based on the "severity of the risk" and the "probability of harm occurring."

The estimated values for each risk item in this device were all within the range of 5 to 9, with the highest being 9. In the evaluation guidelines, a risk estimate of 6 or less is deemed to be sufficiently low that no risk reduction measures are necessary. A risk estimate of 7 to 14 is deemed to be within the acceptable risk level, as "risk reduction is desirable, but if this is not technically or practically realistic, it can be addressed through management measures and warning displays."

Based on these results, it was determined that this device poses as little risk as reasonably practicable (ALARP). ALARP is an international standard principle in safety engineering, and its basis is to "reduce risk as far as reasonably practicable." In the current design, mechanical and electrical safety measures have been implemented to the greatest possible extent, and this device can be evaluated as having risk management based on the ALARP principle.

## 7. Discussion

In this study, to evaluate the movement accuracy and reproducibility of the standing-up support robot we developed, we conducted comparative experiments in which the distance to the target trajectory and the RMSE to the average trajectory of each trial were used as indicators, and the speed and load conditions were changed. As a result, the error between the measured trajectory and the target trajectory was within about 4% of the movement amount under all conditions, and it was confirmed that the robot could reproduce the standing-up movement with a certain degree of accuracy and stability.

Although the control method of this robot is limited to feedforward-type speed control using pulse commands, it is believed that sufficient trajectory tracking was achieved due to the geometric constraints of the link mechanism. In addition, the time-series waveform data of the motor load rate in the five trials was well synchronized, and it is possible that the processing of the motion influences the variation in the trajectory shown in Figures 10 to 12 capture data, especially the detection of the start and end times of the movement and the accuracy of data extraction.

On the other hand, the maximum load in this experiment was limited to 20 kg, and the evaluation of the stability of the operation in a real environment against the movement of the human body and the accompanying large disturbances (e.g., sudden intervention by a caregiver, unexpected physical movement of the user) is a future issue. In addition, this evaluation was performed under conditions where the target trajectory was set in advance, and in order to evaluate the flexibility and adaptability of the system to follow human movements, an experimental design including dynamic tasks and cooperative movements with humans will be necessary. In the future, by realizing a device that can reproduce an individual's natural standing trajectory with high accuracy, it is thought that it will be possible to quantitatively clarify the effects on the support effect of trajectory distortion caused by using an averaged trajectory or a fixed link structure, as well as the discrepancy with the user's original movement.

In the future, with a view to practical application in clinical settings and home care environments, comprehensive verification is required that incorporates individual optimization according to the subject's physical characteristics and needs, as well as subjective evaluation of safety and comfort during actual operation. Through demonstration experiments with subjects with diverse physical characteristics, the development and evaluation of a control method that emphasizes cooperation with humans, and investigations into the subjective comfort and psychological burden of users, it is expected that the system will develop into a highly implementable support technology for practical use.

## 8. Conclusion

In this study, to verify the reproducibility and robustness of the standing assistance robot, we quantitatively evaluated its performance under different speed and weighting conditions, using the distance to the target trajectory and the RMSE to the average trajectory of each trial as evaluation indices. As a result, it was confirmed that the error between the target value and the actual value was sufficiently small at about 4% under all conditions, indicating high accuracy and stability of the movement.

The above results suggest that a non-wearable assistance mechanism utilizing geometric constraints can achieve high trajectory reproducibility while maintaining simple control, and are believed to be an effective guideline for the design of future collaborative standing assistance robots.

## Acknowledgments

This research was partly supported by the Suzuki Foundation 2024 Science and Technology Research Grant (General) and the Osaka Public University 2024 RESPECT Joint Research Grant.

# Tables

Table1: Parameters

| Symbol | Item | Value / Range/Unit |
|---|---|---|
| $x_a, y_a$ | Ankle joint center coordinates | $(0, 0.090)$ m |
| $x_p, y_p$ | Main actuator rotation center coordinates | $(0, 0.270)$ m |
| $A_x$ | Horizontal distance of link $A$  $L_{STL}$ minus horizontal offset | $0.270$ m |
| $A_y$ | Vertical distance of link $A$  Fixed value plus $y_a$ | $0.195+0.090$ m |
| $A$ | Pedestal link $A$  Ankle joint center-main actuator rotation center distance | $\sqrt{A_x^2 + A_y^2}$ |
| $C_x$ | Horizontal distance of link $C$ | $L_{STL} - 0.126$ m |
| $C_y$ | Vertical distance of link $C$ | $H_{SHJ} - 0.033$ m |
| $C$ | Link $C$ knee joint slider center distance | $\sqrt{C_x^2 + C_y^2}$ |
| $D$ | Lower leg link length | $0.390$ m |
| $\theta_{Offset}$ | Angle between thigh and link C | $\tan^{-1}(C_y/C_x)$ |
| $L_{STL}$ | Sitting lateral epicondyle-trochanter horizontal distance  AIST J7 Seated buttock-lateral epicondyle distance (Young's average + Eldery's average)/2 - AIST J9 Seated buttock-trochanter distance (Young's average + Eldery's average)/2 | $0.396$ m |
| $H_{SHJ}$ | Seated trochanter height  AIST I12 (Young's average + Eldery's average)/2 | $0.067$ m |
| $\alpha_a$ | Ankle joint angle | |
| $\alpha_k$ | Knee joint angle | |

# Supplement Materials

## 1. 人体寸法データベースに基いた機構設計

In the mechanical design of this device, we considered the actuator mounting position, which is difficult to adjust due to individual differences, to accommodate the widest possible range of users' heights and to minimize thrust loss. The design conditions, taking into account the constraints of the actuator specifications, were set as follows: stroke within 500 mm, speed within 400 mm/s, thrust 400 N or less for the main actuator, and stroke within 100 mm, speed within 300 mm/s, thrust 360 N or less for the sub-actuator.

In examining the parameters, we used the human body dimension database (1991–1992) of the National Institute of Advanced Industrial Science and Technology (AIST), and created segment length patterns based on the following representative values: minimum value (minimum dimension value in the database), 2.5th percentile, 5th percentile, and 10th percentile for elderly people, average value for elderly people, overall average value including young people and elderly people, average value for young people, maximum value (maximum dimension value in the database).

The actuator mounting position was determined by the designer, without the use of an automatic optimization algorithm. As a result of the evaluation, it was found that short height parameters below the 10th percentile for the elderly population exceeded the sub-actuator's range of motion (100 mm), and therefore it was decided that these users would not be eligible for assistance.

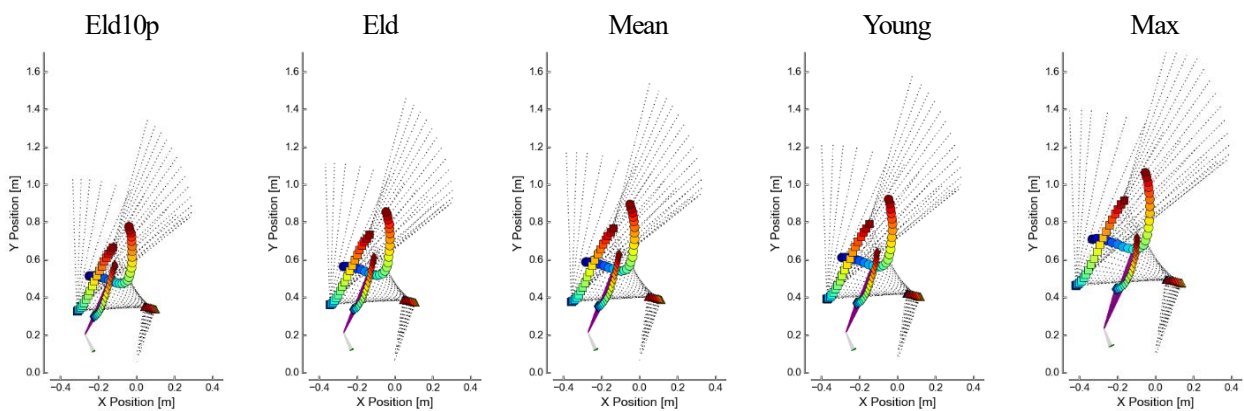

**Fig. S1: Example of trajectory design evaluation by height.**

Using multiple parameter sets based on AIST's human body dimension data, we verified whether they fit within the actuator stroke range. Some of the minimum height groups were not eligible for assistance.

## 2. Behavior of motor load rate and sagittal plane trajectory under 20 kg load condition.

Figure S2 shows the time series of the motor load rate of each actuator (sub, left, right) under the 20 kg load condition. Measurements were performed once each under speed settings of 5 and 10. As with the 0 kg and 10 kg conditions, some actuators had load rates approaching 100% during the standing motion, but it was confirmed that the standing motion was completed normally under all conditions. However, after the motion was completed, that is, when the pulse signal stopped, the sub-actuator load rate was observed to be stuck at 100%, and the motor driver's factory-set safety timer was activated, leading to an automatic stop. The driver's safety setting time was not changed, and measurements under the 20 kg condition were completed after six times. This is because the sub-actuator was continuously loaded at the end state when the seat was at its most forward and upward position. However, in actual use, the user's body is not fixed to the device, and a situation in which such a load is applied when the user completes standing is unlikely to occur. Therefore, this event is judged not to pose a fatal safety risk.

Figure S3 shows the actual trajectory of the robot under the same conditions. As with the 0 kg and 10 kg conditions, the actual trajectory tends to deviate slightly forward from the target trajectory, but the S-shaped trajectory of the hip joint and the circular trajectory of the knee joint are both reproduced well. The reason why the trajectory extends diagonally downward and forward in the final standing position is due to the effect of the free fall caused by the sub-actuator stopping as mentioned above.

The above shows that this device has the performance to stably and safely provide the necessary assistance during the standing movement even under a load of 20 kg. Furthermore, in a separate preliminary experiment, normal standing assistance was provided to a healthy subject weighing 90 kg, and this device is considered to have sufficient load-bearing capacity.

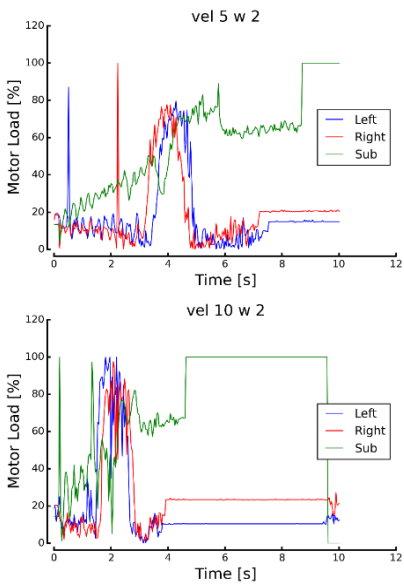

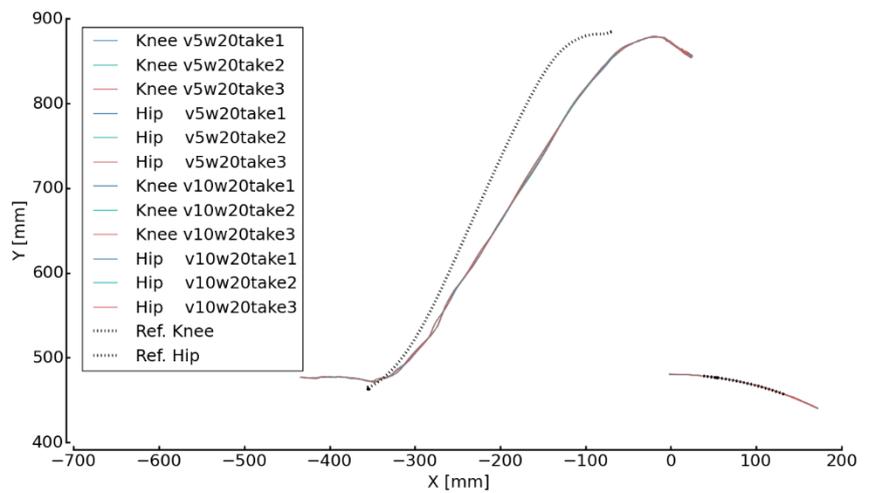

**Fig. S2. Motor load profiles under 20 kg load condition.**
Motor load rates for the sub, left, and right actuators under 20 kg loading during stand-up motion.

**Fig. S3. Measured sagittal trajectories under 20 kg load condition.**
Reproduced trajectories of the hip (S-shaped) and knee (circular arc) under 20 kg loading, plotted in the sagittal plane. While the endpoint shifts slightly forward due to load, overall trajectory reproduction remains accurate. The final drop in hip position reflects passive descent after sub-actuator disengagement.